# Automated Usability Testing: Analysing Asia Web Sites

Rukshan Alexander · Baravalle Andres

**Abstract** Web usability is continuing to be a pressing problem. For number of years researchers have been developed tools for doing automatic web usability testing. This study uses our own PHP, and MySQL based tool AWebHUT: Automated Web Homepage Usability Tester to evaluate web usability of full Dmoz (www.dmoz.org) Asia web sites (45126 on time stamp 2011-12-03 04:12:46 GMT). The tool uses an extensive automated quantitative analysis of XHTML source code of homepages against seventeen organised web usability guidelines. The automated quantitative approach is effective on large scale to achieve better usability. The AWebHUT uses four web usability levels such as N: Neutral, V: Violate, R: Respect, and E: Error to evaluate web usability. The main objective of the study is to produce data which is used to answer research questions, (1) Are there any categories of web sites which have usability problems? Which ones? and (2) Are there any categories in which the usability is typically higher? Why? The findings were indicated that all Asia categories have usability problems. Furthermore, there are four web sites which have highest web usability problem with violation percentage 71. One step further, the Asia category: Weather has highest usability problems with 42.2819 as the average of the violation percentage. The category Weather uses tables and images, considerable amount of those were not satisfying web usability guidelines which relates to tables and images. One step further, the Asia wants to get the same level of usability as North America, Europe, and Australia therefore it is essential to have an automated web usability evaluation in Asia web sites to identify web usability problems which are important for improving Asia web sites.

**Key words** Web-usability · Web-site-evaluation

This research was supported by a research grant from the Vavuniya Campus of the University of Jaffna of Sri Lanka to Rukshan Alexander (UJ/VC/RG/2011/3).

R. Alexander
Faculty of Business Studies, Vavuniya Campus, University of Jaffna, Pampaimadu, Vavuniya, Sri Lanka
e-mail: rukshan@mail.vau.jfn.ac.lk

B. Andres
School of Computing, Information Technology and Engineering (CITE), University of East London, 4-6 University Way, London E16 2RD, UK.
e-mail: a.baravalle@uel.ac.uk

## 1 Introduction

Vast amounts of data are available on web pages, but they can be presented in a variety of ways which will determine the web page tasks, users and environment (Abdulkhair 2004; Zihou 2009; Rukshan and Baravalle 2011). However, a large amount of data is not reachable to all users, due to usability and accessibility problems in web sites (Jasselette 2007). The emergence of this problem led to the launch of construction standards and usability guidelines (Jasselette 2007). A wide variety of usability guidelines and recommendations exist each of them is concerned with how to achieve the user satisfaction and to present a usable web design. Unfortunately, studies carried out that evaluating web sites by reviewing usability guidelines and recommendations reveal that in reality this is difficult to conduct, essentially because of the inappropriate way they are being formulated and structured (Scapin et al. 2000; Ivory and Hearst 2001; Beirekdar et al. 2003; Zihou 2009). In an effort to address this problem, different evaluation techniques have been constructed and used (Vanderdonckt and Berekdar 2005; Rukshan and Baravalle 2011). All these techniques require usability specialists to conduct them or to analyze evaluation results, which consume large amounts of time in addition to other resources especially for very large, continuously growing web sites (Beirekdar et al. 2003; Vanderdonckt and Berekdar 2005; Rukshan and Baravalle 2011). In addition, there is a lack of experts due to an increasing demand of usability analysts (Vanderdonckt and Berekdar 2005). For this reason and others, several automatic evaluation tools were developed to organise web usability guidelines for automation and assist evaluators with guidelines that automatically detect and report ergonomic deviations such as usability, accessibility, etc (Zihou 2009).

Despite a web site being made up of numerous pages, the homepage can be considered to be a true reflection of itself (Rukshan and Baravalle 2011). Therefore it would be sufficient to solely evaluate the homepage of a web site as this page receives the greatest number of visitors (Nielsen and Tahir 2002; Rukshan and Baravalle 2011). Furthermore, all web pages of web sites are created by using XHTML format code which codes for the different characteristics of a web site. By testing through automated analysis of XHTML source code against usability guidelines, it can be efficiently evaluate web pages against web usability problems and produce results of web usability.

The Internet passed 2 billion users in early 2011 and most growth was in Asia (Useit 2012). Asia is warming up to usability testing, and it's about time, says web usability expert Jakob Nielsen (ZDNet 2007). Furthermore, Jakob Nielsen pointed out that the Asia wants to get the same level of usability as North America and Europe (Useit 2012). Therefore, there is a real need to investigate web usability of Asia web sites using a wider, systematic approach. The producing results are important for improving web usability of Asia web sites and for reflection on the underlying information architecture of web sites. Accordingly, this study focuses on two research questions: (1) Are there any categories of web sites which have usability problems? Which ones? (2) Are there any categories in which the usability is typically higher? Why? In part of the research, a tool AWebHUT has been created in PHP (PHP 2009) to perform automated usability testing of homepages by the help of Apache web server and the MySQL (MySQL 2010) database for providing data to answer the above research questions. This research provides an extensive automated quantitative analysis of web usability, using full Dmoz ([www.dmoz.org](www.dmoz.org)) (Benjamin et al. 2006) catalogue of Asia web sites focusing on homepages.

## 2 Review of Relevant Literature Pertaining to the Research

### 2.1 Usability

The term "user friendly" (Bevan et al. 1991; Nielsen 1993) is most important one after user interface increased significantly when the personal computer revolution started (Rukshan and Baravalle 2011). According to ISO9241 usability can be defined as "the effectiveness, efficiency and satisfaction with which specified users achieve specified goals in particular environments", where (Brajnik 2000b):

- Effectiveness refers to the both the completion and precision a consumer/user can success in a desired task/action.
- Efficiency refers to the precision and net use of resources and the extent to which an action is completed.
- Satisfaction refers to a user being content and comfortable with the use of the work system.

### 2.2 Need for Homepage Usability

The usability is today widely recognized as an important requirement for user acceptance. Nowadays electronic commercial web sites are growing rapidly (RedTechnology.com 2010). The homepage's impact on a company's bottom is far greater than simple measure of e-commerce revenues: the homepage is also the company's face to the world. As the homepage is the most important, on any customers' dissatisfaction will make customers to go to other sites (Nielsen and Tahir 2002; Rukshan and Baravalle 2011). Furthermore, while the homepage can be often the customer's first – and possibly last chance to attract customer, rather than the front page of a news paper (Rukshan and Baravalle 2011; Nielsen and Tahir 2002). Therefore, it is important to maintain homepage more usability to users.

2.3 Usability Evaluation

The ISO 13407 (ISO 1999) defines the usability evaluation as follows, measuring the effectiveness, efficiency and satisfaction of the interface usability is called the usability evaluation. Usability evaluation is an important interface design process (Ivory 2001) because of the great benefits of discovering the problems of the design and of allowing a better understanding of the targeted users (Nielsen 1993).

2.4 Usability Evaluation Methods

The usability evaluation methods are classified into two-dimensional approaches, empirical evaluation and analytical evaluation (Parush 2001; Brinck and Hofer 2002). In empirical evaluation, the users are directly involved in the evaluation process to some extent (Nielsen 1993; Brinck and Hofer 2002), whereas in analytical evaluation, various combinations of guidelines, criteria and models are involved in the assessment procedure (Brinck and Hofer 2002). All the usability evaluation methods are based on testing, inquiry, simulation and inspection methods (Abdulkhair 2004).

2.5 Usability Guidelines

To satisfy users by presenting a usable web design, a wide variety of usability guidelines exist, and has been established by different authors (Ivory 2001; Nielsen and Tahir 2002; Rohn et al. 2002; Graham 2003; W3C 2008a; Zihou 2009). Each one of them focuses on how to satisfy users by presenting a usable design and also these guidelines address a wide range of web page design issues starting from the browser's title to the detail of web page features (Rukshan and Baravalle 2011).

According to (Scapin et al. 2000) guidelines are numerous and distributed among different sources: recommendation papers, design standards, style guides that are specific to a particular environment, design guides and algorithms for ergonomic design. Additionally, contradictions exist between guidelines, which might be because of their different ages, the changes in the technology or because they might be appropriate only for specific group of users (Rukshan and Baravalle 2011).

The Web Content Accessibility Guidelines (WCAG 2.0) (W3C 2008a) play an important role in usability guideline because, it yields to apply broadly to more advanced technologies; is easier to use and understand; and is more precisely testable with automated testing and human evaluation (W3C 2008b).

Furthermore while usability implies accessibility (at least when an unconstrained user population is considered), the contrary is not necessarily true. For example, a missing link to the homepage may be a fault affecting usability, while it does not affect accessibility (Brajnik 2000a). According to the Brajnik (2000a) point of view it is possible to categorise guidelines within the following,

1. Consistency of presentation and controls
2. Adequate feedback
3. Natural organization of the information
4. Contextual navigation
5. Efficient navigation
6. Clear and meaningful labels.
7. Robustness
8. Flexibility
9. Functionality
10. Others

Although any of the usability guidelines apply in general, researchers can filter it according to the scope of homepage usability and also have usability guidelines as a checklist when designing homepages (Nielsen and Tahir 2002; Rukshan and Baravalle 2011).

**Table 1** Comparison of classic and automated usability evaluation methods

| Classic Usability Evaluation | Automated Usability Evaluation |
| --- | --- |
| Time consuming | Faster |
| Expensive | Cheaper |
| Require engineers that specialise in usability on call | Automated, remote |
| Small numbers of subjects | Larger numbers of subjects |
| Local users | Intercity, interregional, and international users |
| Controlled Lab | Home or Workplaces |
| Depth | Breadth |
| All (Web site, Software, Web Applications) | Web site, E-Commerce, WWW interfaces |

2.6 Automated Usability Evaluation of Web Sites

Exponential growth of web sites cause web site usability to become evermore significant as the number of users increases (Ivory 2001). Classic usability evaluation methods are expensive, time and resource consuming, could require engineers that specialise in usability (Scholtz et al. 1998). Automated usability evaluation focuses on developing tools and techniques to rapid evaluation tools for larger numbers of subjects that reach a wider audience with intercity, interregional, and international users for usability testing, and tools that have built-in analyses features of automated remotely (Rukshan and Baravalle 2011). Furthermore, the Table 1 compares the classic and automated usability evaluation methods. Different automatic web usability evaluation tools have been established to solve the diversity web interface problems (Abdulkhair 2004). Representative examples of these tools include: WebSat (Scholtz et al. 1998), A-Prompt (A-Prompt 1999), LIFT (Brajnik 2000b), WebTANGO (Ivory 2001; WebTango 2002; Ivory and Hearst 2002), Bobby (Beirekdar et al. 2003), and KWARESMI (Beirekdar et al. 2003). Most of these tools share the same technique: carry out an analysis of the XHTML source code of the web page in context, capture the maximum amount of useful data and identify targeted usability problems (Beirekdar et al. 2003; Zihou 2009; Rukshan and Baravalle 2011).

2.7 Web Usability in Asian

Asia is the world's largest and most populous continent, located primarily in the eastern and northern hemispheres. It covers 8.6% of the Earth's total surface area (or 29.9% of its land area) and with approximately 4 billion people, it hosts 60% of the world's current human population (Exploredia 2011). The Internet passed 2 billion users in early 2011 and most growth was in Asia, which will hit a full billion users in its own right during 2012 (Useit 2012). In other parts of the world have been ahead of Asia with building usability into their web site designing and with hiring usability professionals or experts. In earlier years, Asian countries were more interested in technology for its own sake and built web sites and software that were much too complicated. In the last two year or three, Jakob Nielsen felt a growing Asian interest in web usability. Asian contains 52 countries in its region (CIA 2011). And In order to primarily analyse web usability of Asia web sites, it will be using [www.dmoz.org](www.dmoz.org) to provide a list of categorised web sites that are determined to be from the Asia Region. The ODP (Benjamin et al. 2006) also known as Dmoz is a multilingual open content directory of WWW links. It helps to build the most comprehensive human-reviewed directory of the web. At present 37 countries (as on 2011-12-03 04:12:46 GMT) have been registered as Asia countries in [http://www.dmoz.org/Regional/Asia/](http://www.dmoz.org/Regional/Asia/). The Dmoz Asia has the following 14 categories of web sites which are shown in Table 2.

**Table 2** Dmoz Asia Category (as on 2011-12-03 04:12:46 GMT).

| # | Dmoz Category |
| --- | --- |
| 1 | Arts_and_Entertainment |
| 2 | Business_and_Economy |
| 3 | Education |
| 4 | Government |
| 5 | Guides_and_Directories |
| 6 | Health |
| 7 | Maps_and_Views |
| 8 | News_and_Media |
| 9 | Recreation_and_Sports |
| 10 | Science_and_Environment |
| 11 | Society_and_Culture |
| 12 | Transportation |
| 13 | Travel_and_Tourism |
| 14 | Weather |

## 3 Methodology

### 3.1 Goals

The AWebHUT has been constructed in order to fulfil the following goal for automatic web usability testing of Asia web sites:

- Provide a summarised report about respecting and violation of web usability guidelines of Dmoz Asia web site homepages.
- Provide a customised report for the level of web usability of the Dmoz Asia category web sites which will focus to provide data to answer the following research questions presented earlier.

For achieving the above goals I have sub goals as follows,

- Find and store Asia web sites and related details such as address, sites category path, and site description.
- Organise some basic web usability guidelines for automation.
- Provide web oriented work on XHTML source code of homepages as their code is usable with set of web usability guidelines.
- Provide an automated web usability evaluation task against Asia homepages.
- Provide a customising functionality for evaluating the web usability of the Asia category.
- Provide summarised web usability testing report of web sites with respect of there web usability levels such as violation, respection, neutral, and error for guidelines.

### 3.2 Organising Web Usability Guidelines and Targeting XHTML Elements

One of the main activity of this study is to organise some basic web usability guidelines for automation while, at the same time, test the homepages by using organised web usability guidelines. Several XHTML elements such as tags, attributes, values and labels were used to evaluate a homepage against the targeted web usability guidelines. The web page's Document Object Model DOM (Suhit et al. 2003) structure is used to identify XHTML elements. The following Table 3 indicates the selected web usability guidelines with category for providing automation to evaluate homepages in AWebHUT.

**Table 3** Guidelines used to evaluate Asia web sites by the AWebHUT.

| Guideline Category Serial Number | Guideline Category | Guideline Serial Number | Guideline |
|---|---|---|---|
| 1 | Consistency of presentation and controls | 1.1 | Link label: different links pointing to the same resource should have the same label |

| 2 | Adequate feedback | 2.1 | Freshness: pages should be time - and author-stamped |
|---|---|---|---|
| 3 | Contextual navigation | 3.1 | NOFRAMES validity: NOFRAMES should be present and it should contain equivalent navigation options |
| | | 3.2 | Link to home: each page should contain a link to the homepage |
| | | 3.3 | Frame titles: frames should set the "title" attribute |
| 4 | Efficient navigation | 4.1 | Table coding: table components should have explicit width and height |
| | | 4.2 | Image coding: images should also have explicit width and height |
| 5 | Clear and meaningful labels | 5.1 | Explicit mailto addresses: labels of "mailto" links should contain the actual email address |
| | | 5.2 | Missing page title: pages should have a title |
| | | 5.3 | Table headers: tables should have headers |
| 6 | Robustness | 6.1 | Link targets: avoid "_blank" target in frames |
| | | 6.2 | Portable font-faces: standard font faces should be used in addition to desired ones |
| 7 | Flexibility | 7.1 | Image ALT: images should have alternative textual descriptions |
| | | 7.2 | Frames resizing: relative sizes should be used |
| 8 | Support of users goals | 8.1 | Form coding: forms should have "submit", "reset" buttons |
| 9 | Other | 9.1 | Keywords/description: pages should have appropriate META information to be searchable by search engines |
| | | 9.2 | Marquee,blink: avoid animated features |

The following is one of the sample algorithm to evaluate the Guideline 7.1: Image ALT: images should have alternative textual description.

```
Being
    If (there is no image tag) then
        Return the value as neutral
    Else
        Do
        Begin
            If (alt attribute not exists OR the value of alt equals "") then
                Return the value as violation
        While for each image tag
        Return the value as respecting
    End if
End
```

3.3 Tool Architecture

In order to reach the goals stated, the AWebHUT activities can be characterized by the following sequence of activities:

1. Filter and store Asia web sites
2. Page parsing and data generation
3. Web usability problem detection
4. Visualization of results as summarised report

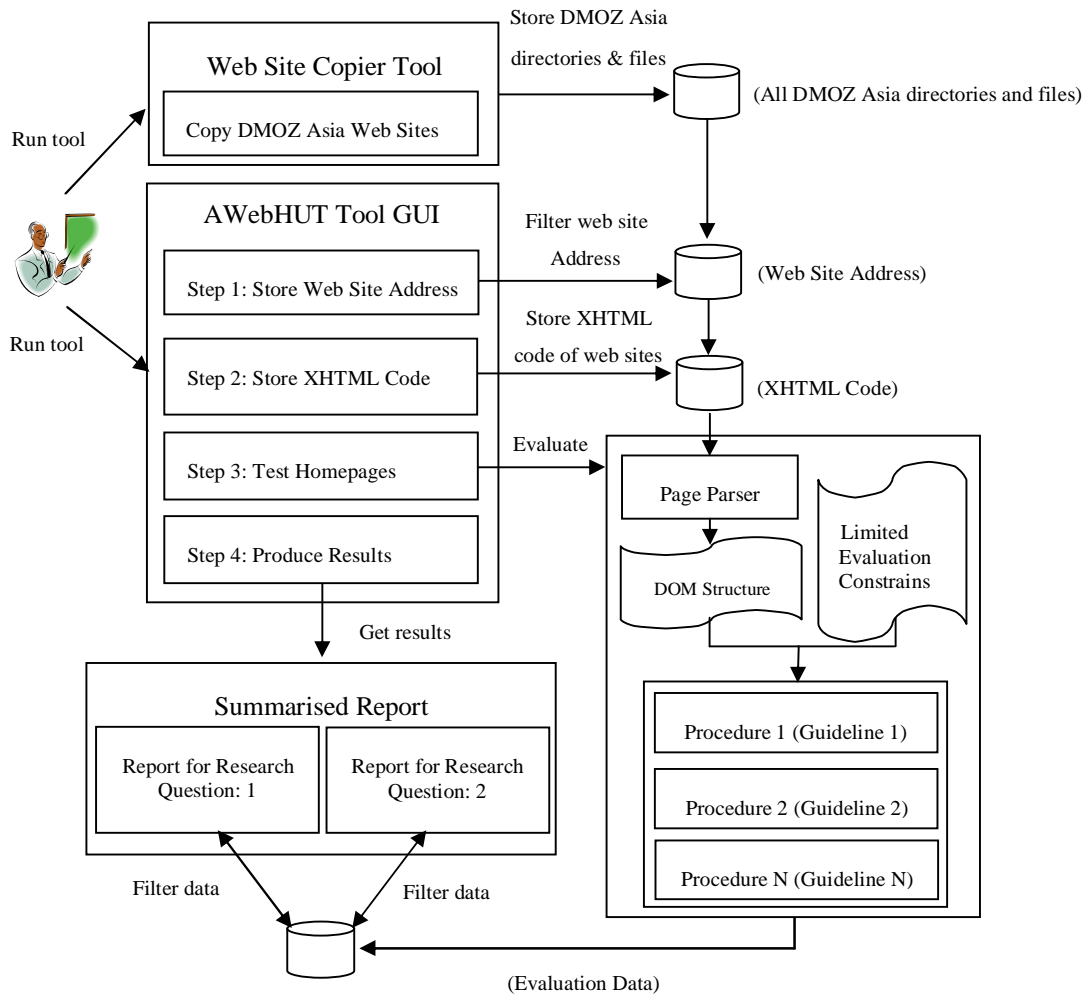

**Figure 1** AWebHUT tool architecture

The AWebHUT uses XHTML source code analysis techniques. The XHTML source code of homepages is analyzed at design/definition time. Figure 1 illustrates the architecture of the AWebHUT. The AWebHUT is working with the support of a web site copier tool. The Figure 1 shows how the AWebHUT and the web site copier tool will work in conjunction with each other to fulfil our goal.

To evaluate web usability, the evaluator wants to run one of the web site copier tools which will store all Dmoz Asia directories and its files of the Asia region. Then, step 1 of AwebHUT want to perform for storing Asia web site address into database. The step 1 recursively scan through all stored Dmoz Asia directory's index.html files for scanning and filtering Asia web site address then store it into database. Then, step 2 of AWebHUT want to perform for storing XHTML source code of web sites. It catches the web site address from the database and passes it to get its XHTML source code structure of the homepages of the web sites then store the XHTML source code into database. Then, step 3 of AWebHUT want to perform for testing web usability of web pages. It gets the XHTML source code structure of homepages; it performs web usability guideline evaluation one by one by calling procedures in sequential manner for which researchers have developed a separate algorithm in the AwebHUT. The sequential manner of evaluation is very suitable for the AWebHUT architecture because each web usability guideline is represented and implemented with its own straightforward procedure which yields to gain efficient and fast evaluation process. For each web usability guideline evaluation for a homepage several metrics are generated. Finally, the step 3 of AWebHUT want to perform for producing summarized report with customisation capabilities about the web usability testing of Dmoz Asia web sites.

3.4 Step 1: Store Web Sites

In the Dmoz Asia category, all web site links has been bounded with in the tag of "<ul class= 'directory-url'>" in its index.html file. It is possible to filter web site address and relevant details by analysing XHTML source code of index.html file of the category. The filtered data are stored into database.

The following is the algorithm to filter the web site address in the AWebHUT tool.

Being
   Do
   Begin
      If (variable v's attribute 'class' value equals 'directory-url') then
         Do
         Begin
            Store variable u's children(0)'s attribute href's value and other necessary details into database.
         While for each variable v's tag 'li' as u
      End if
   While each page's DOM structure, find tag 'ul' as variable v,
End

3.5 Step 2: Store XHTML Code

To provide data for the web usability guideline evaluation process, the XHTML source code of Dmoz Asia web sites homepages were stored. The AWebHUT's XHTML code parsing strategy is based on scanning the whole web page and extracting the tags, attributes, values, and labels that we used when evaluating our organised web usability guidelines in term of procedures. For this purpose the DOM is used. Researchers assume that all homepages have no XHTML code syntax errors, areas of non-conformance to coding standards and browser compatibility problems.

3.6 Step 3: Test Homepages

To evaluate the web usability of homepage of web site, researchers use the evaluation algorithm that was defined to determine the order of checking each web usability guidelines. Furthermore, each web usability guidelines have own its algorithm. Checking an evaluation by web usability guideline is done by checking its instances that were selected in the DOM parsed page to see if they respect or not for the evaluation condition that was associated with the web usability guideline.

3.7 Step 4: Produce Results

The AWebHUT tool produces reports as summarised report. In the summarised report, four web usability levels are used to measure the web usability of homepages are as follows,

1. N – Neutral: The homepage is being neutral for the web usability guideline.
2. V – Violating: The homepage is violating the web usability guideline.
3. R – Respecting: The homepage is respecting the web usability guideline.
4. E – Error: The homepage has considered as error because of one or more of the following reasons,

a. The web address is redirected to another site.

   b. Threads are found in the web sites such as, Malware Objects, Banners, Phishing Sites, Spam, Malicious Scripts, and Network Attacks

   c. Client or server side network Internet connection fails during the web   usability evaluation

**4 Assumptions and Limitations**

It should be emphasised that part of the study addresses selected web usability guidelines based on categorise identified above. The determination all involving XHTML elements to perform web usability guideline review is a crucial step for the identification of web metrics (Abdulkhair 2004; Zihou 2009; Rukshan and Baravalle 2011). To perform the web usability evaluation, the authors assume there are no syntax errors in the XHTML coding of the homepages and also the authors did not consider the XHTML page layout effects such as Java scripts, CSS, PHP. Furthermore, most Asian countries have more than one language that is natively spoken (CIA, 2011). The authors assume that all Dmoz Asia web pages are in English, and these guidelines can sufficiently evaluate English web pages.

 During the automated web usability evaluation several errors were occurred in all steps done by AWebHUT. The Table 4 shows the possible reasons for errors in each step during the automated web usability evaluation.

**Table 4** Errors in each step during the automated web usability evaluation

| Process | Results Description | Total Results |
| --- | --- | --- |
| Initial Dmoz Asia Web Sites | Total number of web sites for the automated web usability evaluation | 45126 |
| Step 1: Store Web Sites | Errors: Total errors of Step 1<br>Error Reason:<br>• Evaluator's mistake. | 1 |
| | Success: Total stored web sites | 45125 |
| Step 2: Store XHTML Code | Errors:  Total errors of Step 2<br>Error Reason:<br>• Too much of time requires for storing the source code of web site.<br>• No response from server of the web site.<br>• Server of the web site blocks the request. | 2956 |
| | Success: Total stored XHTML source code | 42169 |
| Step 3: Test Homepages | Errors:  Total errors of Step 3<br>Error Reason:<br>• Some part of the web site XHTML code blocks the evaluation of that web site. | 337 |
| | Success: Total tested homepages | 41832 |
| Step 4: Produce Results | Errors:  Total errors of Step 4 | 0 |
| | Success: Total produced results | 41832 |

**5. Findings**

The results of this study are organized by the two research questions presented earlier. The Figure 2 was computed by the AWebHUT which provides data to answer the research question 1. According, in Figure 2 four web sites have more usability problems with the high violation percentage of 71 and which are listed in Table 5.

| Web Site Address | Category Path | 1.1 | 2.1 | 3.1 | 3.2 | 3.3 | 4.1 | 4.2 | 5.1 | 5.2 | 5.3 | 6.1 | 6.2 | 7.1 | 7.2 | 8.1 | 9.1 | 9.2 | % of V |
|---|---|---|---|---|---|---|---|---|---|---|---|---|---|---|---|---|---|---|---|
| http://manresa87.freeservers.com/ | www.dmoz.org/Regional/Asia/Philippines/Education/Reunions_and_Alumni/ | V | V | N | V | N | V | V | R | V | V | N | V | V | N | V | V | V | 71 |
| http://ddws.nic.in/ | www.dmoz.org/Regional/Asia/India/Government/Agriculture_and_Rural_Affairs/ | V | V | N | V | N | V | V | R | V | V | N | V | V | N | V | V | V | 71 |
| http://2coolindore.8k.com/ | www.dmoz.org/Regional/Asia/India/Madhya_Pradesh/Localities/Indore/Guides_and_Directories/ | V | V | N | V | N | V | V | R | V | V | N | V | V | N | V | V | V | 71 |
| http://skydivemalaysia.tripod.com/ | www.dmoz.org/Regional/Asia/Malaysia/Recreation_and_Sports/ | V | V | V | V | V | V | V | R | R | V | R | V | V | R | V | V | R | 71 |
| http://nefin.org.np/ | www.dmoz.org/Regional/Asia/Nepal/Society_and_Culture/Ethnicity/ | V | V | N | V | N | V | V | R | R | V | N | V | V | N | V | V | V | 65 |
| http://www.gorkhapatra.org.np/ | www.dmoz.org/Regional/Asia/Nepal/News_and_Media/Online_News/ | V | V | N | V | N | V | V | R | R | V | N | V | V | N | V | V | V | 65 |
| http://beekeepingshop.50megs.com/ | www.dmoz.org/Regional/Asia/Nepal/Business_and_Economy/Food_and_Related_Products/ | V | V | N | V | N | V | V | R | R | V | N | V | V | N | V | V | V | 65 |

**Figure 2** List of web usability evaluation results by descending order of percentage of violation of web sites (screenshot computed by AWebHUT).

All the four web sites which are listed in Table 5 have been violated web usability guidelines of the following, 1.1, 2.1, 3.2, 4.1, 4.2, 5.3, 6.2, 7.1, 8.1, and 9.1. Then, the Table 6 which was computed by the AWebHUT which provides data to answer the research question 2. According, in Table 6 the Asia category: "Weather" has web usability problem which is typically higher with the % average of the violation: 42.2819 while the mean of the % of average of the violation: 37.76. The results for the remaining thirteen measures considered in this study appear to offer more substantial opportunity to improve web usability in its category.

**Table 5** Web sites which have more usability problems with high violation percentage of 71.

| # | Category Path | Web Site Address |
|---|---|---|
| 1 | www.dmoz.org/Regional/Asia/Philippines/Education/Reunions_and_Alumni/ | http://manresa87.freeservers.com/ |
| 2 | www.dmoz.org/Regional/Asia/India/Government/Agriculture_and_Rural_Affairs/ | http://ddws.nic.in/ |
| 3 | www.dmoz.org/Regional/Asia/India/Madhya_Pradesh/Localities/Indore/Guides_and_Directories/ | http://2coolindore.8k.com/ |
| 4 | www.dmoz.org/Regional/Asia/Malaysia/Recreation_and_Sports/ | http://skydivemalaysia.tripod.com/ |

**Table 6** List of filtered evaluation result for Asia category web usability test

| S.No | Dmoz Category | Total Occurrences | E* | Taken for the Web Usability Evaluation | Sum of the Violation Percentage | Average of the Violation Percentage |
|---|---|---|---|---|---|---|
| 1 | Weather | 188 | 25 | 163 | 7949 | 42.2819 |
| 2 | Guides_and_Directories | 692 | 68 | 624 | 28047 | 40.5303 |
| 3 | News_and_Media | 1060 | 108 | 952 | 41240 | 38.9057 |
| 4 | Society_and_Culture | 4662 | 301 | 4361 | 178869 | 38.3674 |
| 5 | Education | 2068 | 192 | 1876 | 78312 | 37.8685 |
| 6 | Recreation_and_Sports | 527 | 33 | 494 | 19787 | 37.5465 |
| 7 | Travel_and_Tourism | 4923 | 295 | 4628 | 184783 | 37.5346 |
| 8 | Maps_and_Views | 419 | 38 | 381 | 15542 | 37.0931 |
| 9 | Arts_and_Entertainment | 2364 | 144 | 2220 | 87621 | 37.0647 |
| 10 | Science_and_Environment | 539 | 51 | 488 | 19920 | 36.9573 |
| 11 | Health | 1209 | 101 | 1108 | 44029 | 36.4177 |
| 12 | Government | 2418 | 327 | 2091 | 87830 | 36.3234 |
| 13 | Transportation | 371 | 32 | 339 | 13393 | 36.0997 |
| 14 | Business_and_Economy | 16887 | 1182 | 15705 | 602354 | 35.6697 |

Note. E*: Number of web sites which have web usability level: E (Error) in its all web usability guidelines

The Figure 3 shows why the Asia category: Weather has more web usability problems.

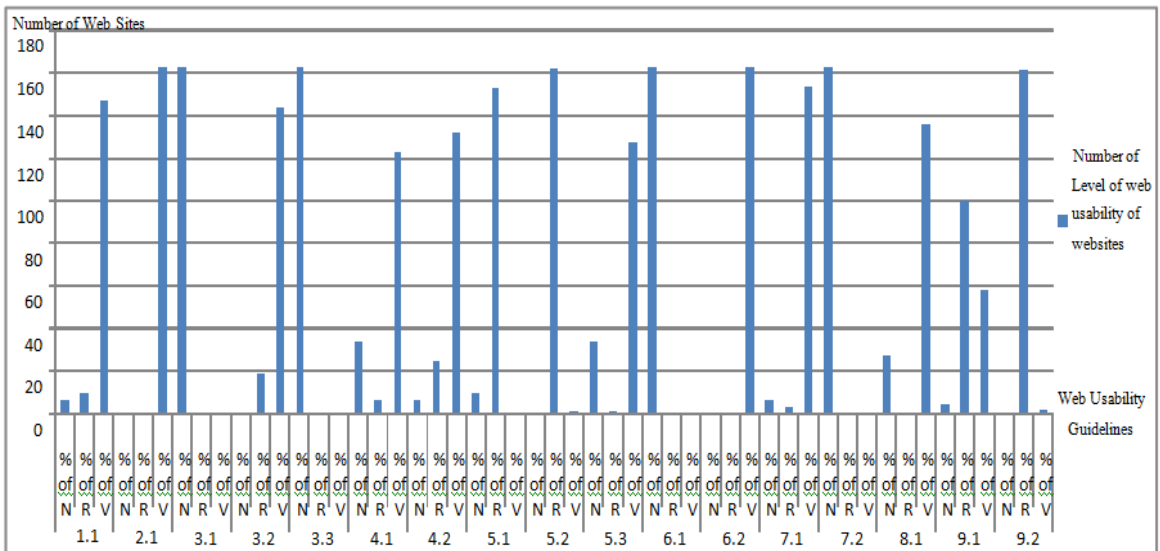

**Figure 3** Reason of why the Asia Category: Weather has web usability problem which is typically higher.

### 6. Conclusions and Recommendations

The users are attracted to WWW services in the present world. Obviously, there is a need to increase the web's usability and accessibility. The WWW vendors, developers, and evaluators spent large amount of time to increase the usability of web. Our research proposed an automated quantitative web usability evaluation method which is suitable for vast and continuously growing web sites. The AWebHUT evaluation method increases the number of usability terms evaluating the system, and increases the number of study participants. It provides the highest level of automation and requires no user testing or informal use.

As, Jakob Nielsen predicted (2012 January) that the web market will continue stronger growth in Asia and slower growth in the almost-fully saturated markets of North America, Europe, and Australia (Useit 2012). It is essential to have an automated web usability evaluation in Asia web sites to identify the web usability problems. Our studies shows that the Asia has web usability problems in there web sites. Furthermore, the category Asia category Weather has highest web usability problem. The evidence is the summarised report which was automatically generated by AWebHUT. The category Weather uses tables and images, considerable amount of those were not satisfying web usability guidelines which relates to tables and images. Furthermore, standard font faces (portable font-faces) were not used in web sites in the category Weather. One step further, the study demonstrated that it is possible to evaluate all Dmoz catalogues of Asia web sites with an extensive quantitative analysis of web usability, focusing on the homepage which is fast and cost less by AWebHUT.

Evaluation result has been stored which can be used for future analysis. Furthermore, the AWebHUT is dependant on another tool of web site copier. It wants to have its own functionality to evaluate any part of the Dmoz directory with out depend on other tools in future. In addition the authors will use information from applying the tool to different web sites to identify new functionality as well as identify limitations of the tool. Future plans include the ability to look at an entire web site at one time, as opposed to the only homepage at a time currently. Then, it yields to analyse for guidelines that cover inter pages relations or link-structure issues can be possible after validating our results in homepage evaluation and also most of the automated tools are fail to evaluate the guideline category of natural organization of the information (Brajnik 2000b). It want to extend the AWebHUT with the colour related usability evaluation will give uniqueness to tool because it is obvious that in order to develop an impartial and comprehensive usability evaluation tool that can assess every aspect within the web page, more usability guidelines should be included. Furthermore, by considering XHTML code syntax errors, areas of non-conformance, layout effects such as Java scripts, CSS, PHP problems then automatically correct problems are also another direction of future work.

**7. References**

A-Prompt (1999). Web accessibility verifier. [online] Available at: http://aprompt.snow.utoronto.ca/. (Accessed: 10/ 04/ 2010).

Abdulkhair, M. (2004). A multilingual automated web usability evaluation agent. *Ph.D. thesis*. The Department of Computer Science, The University of Sheffield.

Beirekdar, A., Vanderdonckt, J., & Noirhomme, M. (2003). KWARESMI – Knowledge based web automated evaluation tool with reconfigurable guidelines optimization. in C, Stephanidis (ed.) *Proceeding of 2$^{nd}$ Int. Conf. on Universal Access in Human – Computer Interaction UAHCT'2003.* Mahwah: Lawrence Erlbaum Associates, 4. pp. 1504 – 1508.

Benjamin, M. G., Erin, M. T., Poh, C. T., Marlene, S., Gurpreet, K. S., John, G., Elena, B. O., & Mark, D.W. (2006). Fast, cheap and out of control: A zero curation model for ontology development. *Pacific Symposium on Biocomputing*, *11*, 128-139.

Bevan, N., Kirakowski, J. & Maissel, J. (1991). What is usability?. *Proceedings of the 4th International Conference on Human Computer Interaction.* Stuttgart. pp. 651-655.

Brajnik, G. (2000a). Automatic web usability evaluation: what needs to be done?. *Proc. of 6th Conf. on Human Factors and the Web HFWeb.*

Brajnik, G. (2000b). Automatic web usability evaluation: where is the limit?. Kortum, Ph., Kudzinger, E. (eds.): *Proc. of 6th Conf. on Human Factors and the Web HFWeb.*

Brinck, T., & Hofer, T. (2002). Automatically evaluating the usability of web sites. *Human Factors in Computing Systems*, Minneapolis, Minnesota, ACM, Press: 906 - 907.

CIA (2011). The world factbook. [online] Available at: https://www.cia.gov/library/publications/the-world-factbook/. (Accessed: 10/09/2011).

Exploredia (2011). Population of Asia 2011. [online] Available at: http://exploredia.com/population-of-asia-2011/. (Accessed: 08/08/2011)

Graham, I. (2003). A pattern language for web usability. London: Addison - Wesley.

ISO (1999). ISO 13407:1999. [online] Available at: http://www.iso.org/iso/catalogue_detail.htm?csnumber=21197. (Accessed: 05/ 04/ 2010)

Ivory, M. (2001). An empirical foundation for automated web interface evaluation. *Ph.D. thesis*. Computer Science. Berkeley, University of California at Berkeley.

Ivory, M., & Hearst, M. (2001). State of art in automating usability evaluation of user interfaces. *ACM computing surveys*. 33 (4), pp. 1- 47.

Ivory, M., & Hearst, M. (2002). Improving design. *IEEE Internet Computing*. pp. 56-63.

Jasselette, A., Keita, M., Noirhomme, M., Randolet, F., Vanderdonckt, J., Brussel, C., & Grolaux, D. (2007). Automated repair tool for usability and accessibility of web sites. Calvary, G. et al. (eds.). *Computer-Aided Design of User Interfaces*. Springer 2007, pp. 261- 272.

MySQL (2010). Mysql Database Products. [online] Available at: http://www.mysql.com/products/. (Accessed: 15/03/2010).

Nielsen, J. (1993). Usability Engineering. San Francisco, California: Morgan Kaufmann Publishers, Inc.

Nielsen, J., & Tahir, M. (2002). Homepage usability 50 websites deconstructed. Indianapolis, USA: New Rider publishing.

Parush, A. (2001). Usability design and testing. *ACM Interactions*, 8(5).

PHP (2009). What is PHP?. [online] Available at: http://php.net/index.php .(Accessed: 05/03/2009).

Rohn, J., Spool, J., & Ektare, M. (2002). Usability in practice: alternatives to formative evaluations - evolution and revolution. *Human Factors in Computing Systems,* Minneapolis, Minnesota, ACM Press. pp. 891-897.

Rukshan, A., & Baravalle, A. (2011). A quantitative approach to usability evaluation of web sites. *Proceedings of Advances in Computing Technology*, London, United Kingdom, 2011.

Scapin, L., Leulier, C., Vanderdonckt, J., Mariage, C., Bastien, C., Farenc, P., & Palanque, R. (2000). A framework for organizing web usability guidelines. In Kortum, Ph., Kudzinger, E. (eds.) *Proc. of 6th Conf. on Human Factors and the Web HFWeb'2000.* Austin: University of Texas, Austin. Accessible at: http://www.isys.ucl.ac.be/bchi/publications/2000/Scapin-HFWeb2000.htm

Scholtz, J., & Laskowski, S. (1998). Developing usability tools and techniques for designing and testing web sites. in *Proceedings of the 4th Conference on Human Factors & the Web HFWeb'98*. Basking Ridge.

Suhit, G., Gail, K., David, N., & Peter, G. (2003). DOM-based content extraction of HTML documents. *Proceedings of the 12th international conference on World Wide Web*, 207 - 214

Useit (2012). Internet Growth: Update January 2012, [online] Available at: http://www.useit.com/alertbox/internet_growth.html. (Accessed: 05/02/12).

Vanderdonckt, J. & Beirekdar, A. (2005). Automated web evaluation by guideline review. *Journal of Web Engineering*. 4 (2), pp.102-117.

W3C (2008a). Web content accessibility guidelines (WCAG) 2.0. [online] Available at: http://www.w3.org/TR/WCAG20/. (Accessed: 20/ 03/ 2010).

W3C (2008b), Web Content Accessibility Guidelines (WCAG) Overview. [online] Available at: http://www.w3.org/WAI/intro/wcag.php. (Accessed: 30/ 03/ 2010).

WebTango (2002). Web Tango: Automating evaluation. [online] Available at: http://webtango.berkeley.edu/. (Accessed: 20/ 04/ 2010).

ZDNet (2007), Usability makes business sense. [online] Available at:

http://www.zdnetasia.com/usability-makes-business-sense-61983822.htm. (Accessed: 05/03/2010).